\documentstyle[prb,aps,multicol,epsf]{revtex}
\renewcommand{\narrowtext}{\begin{multicols}{2}
\global\columnwidth20.5pc}
\renewcommand{\widetext}{\end{multicols}
\global\columnwidth42.5pc} \multicolsep = 8pt plus 4pt minus 3pt

\input{psfig.sty}

\begin{document}

\draft

\title{Kondo effect in multielectron quantum dots 
at high magnetic fields.}

\author{C. Tejedor$^1$ and L. Mart\'{\i}n-Moreno$^2$}

\address{$^1$Departamento de F\'{\i}sica Te\'orica de la Materia
Condensada, Universidad Aut\'onoma de Madrid,  28049 Madrid, Spain.}

\address{$^2$Departamento de F\'{\i}sica de la Materia Condensada, 
ICMA(CSIC), Universidad de Zaragoza, Zaragoza 50015, Spain.} 


\maketitle

\begin{abstract}
We present a general description of low temperature transport through 
a quantum dot with any number of electrons at filling factor $1<\nu <2$.
We provide a general description of a novel Kondo effect which is 
turned on by application of an appropriate magnetic field.
The spin-flip scattering of carriers by the quantum dot only involves two 
states of the scatterer which may have a large spin. 
This process is described by spin-flip Hubbard operators, which change
the angular momentum, leading to a Kondo Hamiltonian. 
We obtain antiferromagnetic exchange couplings depending on tunneling amplitudes
and correlation effects.
Since Kondo temperature has an exponential dependence on exchange couplings, 
quantitative variations of the parameters in different regimes have 
important experimental consequences. In particular, we discuss the 
{\it chess board} aspect of the experimental conductance when represented in 
a grey scale as a function of both the magnetic field and the 
gate potential affecting the quantum dot. 
\end{abstract}


\narrowtext

\section{Introduction}
Kondo effect has been recently observed in quantum dots (QD) coupled to leads 
by analyzing the temperature dependence of the conductance 
\cite{Goldhaber,Cronenwett,Simmel}. Kondo physics has been mainly 
detected for an odd number of electrons $N$ in the QD while for even $N$, 
the temperature dependence of the conductance is usually that 
corresponding to a normal system. This is interpreted in terms of 
the formation of singlets by pairs of electrons so that only when $N$ is  
odd a single electron remains unpaired. 
This last electron, in a state doubly degenerate by spin, is responsible 
for the Kondo behavior proposed long time ago
\cite{Glazman,Ng,Kawabata,Hershfield,Meir,Levi,Matveev}. 
However, some recent experiments
\cite{Sttut,Delft,Delft2} suggests a more complicated situation because Kondo 
effect is also observed for even $N$ in some situations. 
A particularly interesting experimental feature is the appearance, for 
a given $N$, of alternating high and low conductance valleys as a function 
of an external magnetic field $B$\cite{Sttut,Delft2}. When $N$ is varied in 
$\pm 1$, the high and low conductance valleys are interchanged. Therefore, the 
representation of the conductance (in a grey scale) as a function of both $B$ 
and a gate potential which allows to vary $N$, takes 
the aspect of a {\it chess board}\cite{Sttut,Delft2}.

Three models \cite{Pustilnik,Eto,Giuliano} have been already proposed for 
understanding how Kondo effect might be possible for states of two electrons. 
In these approaches, the authors consider the situation of double degeneracy 
that a magnetic field can create between the singlet ($S_z=0$) and the triplet 
($S_z=1$) states of the electronic pair. However, as shown below, 
the actual situation is, in general, rather more complicated due to 
correlation and tunneling amplitude effects.

In this paper we present a general description of the Kondo effect 
in a QD in the presence of a high magnetic field such that the 
filling factor is $1<\nu <2$. This is a regime in which experiments 
clearly show Kondo effect\cite{Sttut}. The description is valid for 
any number of electrons (even or odd without restriction to 1 or 2) 
and any value of the QD spin $S_z$ (not restricted to 0, 1 for even $N$
or 1/2 for odd $N$).
Instead of describing spin-flip scattering in terms of spin-ladder operators 
$S^{(\pm)}$, we find a set of spin-flip 
Hubbard operators describing a collective spin effect of all the $N$ electrons 
contained in the QD. Despite both $N$ and $S_z$ 
can be very large, the scattering of carriers only produces transitions 
between two many-body states of the QD with $S_z$ differing in $\pm 1$.
These many-body states are well characterized theoretically in the regime 
$1<\nu<2$, and we find that the spin-flip process can be described by a 
Kondo Hamiltonian with antiferromagnetic couplings which depend on both 
tunneling amplitudes and correlation effects. 
As a consequence, the system presents a Kondo behavior with a Kondo 
temperature $T_K$ which we analyze in various limiting cases. 

In section II we discuss the ground state (GS) of an isolated QD in the regime 
$1<\nu<2$. in Section III an effective Kondo Hamiltonian is obtained, by 
means of a scattering description of tunneling through the QD. Section IV 
contains a discussion on the exchange couplings and the Kondo temperature in 
the regime $1<\nu<2$. Section V is devoted to the explanation of the 
{\it chess board} aspect of the experimental conductance. 
A summary is given in section VI.

\section{QD spectrum}
We consider $N$ electrons in the presence of a magnetic field and confined in a 
QD coupled to leads. The Hamiltonian is 
\begin{eqnarray}
H=H_{QD}+H_{L}+H_{TUN} . 
\label{hamilt}
\end{eqnarray} 
Within a lowest Landau level approach, an isolated parabolic QD with 
magnetic field and interaction between the electrons is described by 
(hereafter we take $\hbar =1$)
\begin{eqnarray}
& H_{QD} & = \frac {N\Omega}{2}+
\frac{(\Omega-\omega _{c})M}{2} + g \mu_B BS_z + 
\nonumber \\ 
& & 
\frac{1}{2} \sum_{m_i,\sigma _i} V_{m_1m_2m_3m_4}
d^{\dagger}_{m_1,\sigma_1} d^{\dagger}_{m_2,\sigma_2}
d_{m_3,\sigma_3}d_{m_4,\sigma_4} .
\end{eqnarray}
The first two terms describe a single particle contribution depending on  
both the QD confinement $\omega _0$ and cyclotron $\omega _c$ frequencies
through $\Omega=\sqrt{\omega _c^2+4\omega _0^2}$. $M$ is the third component of 
the total angular momentum.
The third term is the Zeeman energy, depending on the Land\'e $g$-factor.
In the last term, describing electron-electron repulsion,  
$d^\dagger _{m,\sigma}$ creates an electron with angular
momentum $m$ and spin $\sigma $ in the QD. The Coulomb interaction matrix 
elements $V_{m_1m_2m_3m_4}$ have a typical energy scale 
$e^2/\varepsilon l_B$, where $\varepsilon $ is the dielectric constant and 
$l_B=1/\sqrt{m\Omega}$ the magnetic length. These interactions decrease with 
the increasing width $w$ of the quantum well in which the QD has been built up. 

$H_{L}=\sum _{k,\sigma} \varepsilon _{k,\sigma} c^\dagger_{k,\sigma} c_{k,\sigma}$
describes, in a single-particle approach, the leads having electrons 
with quantum numbers $k,\sigma$ 
occupying states up to the Fermi energy $\varepsilon _F$. 

A crucial role is going to be played by the tunneling part of the Hamiltonian 
\begin{eqnarray} 
H_{TUN}=\sum_{k,m,\sigma}V_m \left( d_{m,\sigma} ^\dagger c_{k,\sigma}+
c ^\dagger _{k,\sigma} d_{m,\sigma} \right)
\label{htun}
\end{eqnarray}
in which we neglect any dependence on $k$ of the tunneling amplitudes $V_{m}$ 
(taken as real positive) but we retain the dependence on $m$ because 
is going to produce physical consequences.

\begin{figure}
\psfig{figure=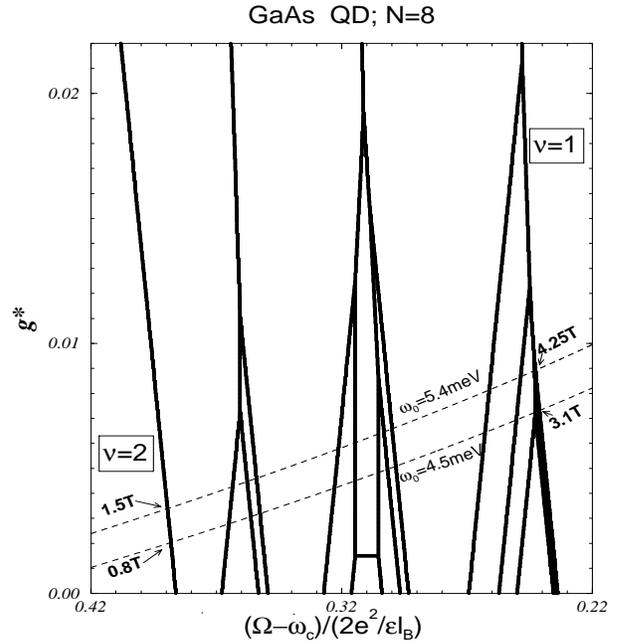,height=9.0cm,width=8.0cm}
\caption{Phase diagram of the possible GS of a GaAs QD with 8 electrons 
at $2>\nu>1$. The axes contain kinetic 
$(\Omega-\omega _c)/(2 e^2/\varepsilon l_B)$ and Zeeman 
$g^* = g\mu _B B/(e^2/\varepsilon l_B)$ contributions to the energy.
The (upper) wider regions correspond to compact 
states from $|C_4^4\rangle $ ($\nu =2$) at the left
to $|C_8^0\rangle $ ($\nu =1$) at the right. 
The (lower) narrower regions correspond to different skyrmion-like states. 
Dashed lines depict, for two different values of the QD 
confinement frequency $\omega _0$, the evolution of the GS within the range 
of magnetic fields given at the edges of the lines.}
\label{fig1}
\end{figure}

\begin{figure}
\psfig{figure=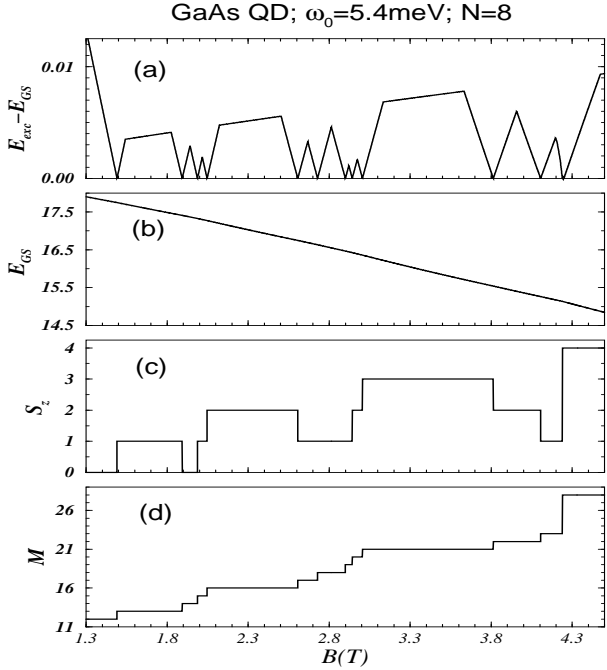,height=9.0cm,width=8.0cm}
\caption{Evolution with the magnetic field $B$ of different magnitudes for 
the GS of a GaAs QD with $N=8$ and $\omega_0=5.4$meV.
Energies are measured in units of $e^2/\varepsilon l_B$. The left 
side of the figure corresponds to $\nu=2$ and the right side to $\nu =1$.}
\label{fig2}
\end{figure}

$H_{QD}$ can be numerically diagonalized for a significantly broad 
range of $N$, as discussed in many theoretical papers\cite{note1}. 
An example (similar results are obtained for an odd number of electrons) 
is shown in Fig. \ref{fig1} and \ref{fig2}, for $N=8$, which 
is the regime in which some experiments \cite{Sttut} have been performed,
although some others \cite{Delft2} involve a larger number of electrons. 
In order to have access to all the possibilities for the GS, it is 
convenient to discuss a case with strong interaction effects, so that 
we present results for a QD with $w=0$. The case of an arbitrary finite 
width $w$ can be qualitatively discussed from these results: correlated 
GS tend to disappear when $w$ increases.
Fig. \ref{fig1} shows the phase diagram of the possible GS 
between $\nu =1$ and $\nu =2$. The wider regions of the upper part of the 
phase diagram correspond to compact states which are favored by large  
Zeeman coupling
\begin{eqnarray}
|C^K_{N-K} \rangle =\prod _{m=0} ^{K-1} d^\dagger _ {m,\downarrow} 
\prod _{m=0} ^{N-K-1} d^\dagger _{m,\uparrow} |0 \rangle 
\end{eqnarray}
where $|0 \rangle$ is the vacuum state.
In going from left to right, one finds successively the states 
$|C_4^4\rangle $ ($\nu =2$), $|C_5^3\rangle $, $|C_6^2\rangle $, 
$|C_7^1\rangle $ and $|C_8^0\rangle $ ($\nu =1$).
Strong interaction effects are manifested in the appearance, lower in the 
phase diagram, of the narrower regions corresponding to different 
skyrmion-like states \cite{Oaknin1,Oaknin2} 
\begin{eqnarray}
|SK^{P}_{N,K,\pm} \rangle={\cal N}_{P,\pm} \left(\Lambda_{\pm 1}^\dagger \right) 
^P|C^{K}_{N-K}\rangle
\end{eqnarray}
of topological charge 1. ${\cal N}_{P,\pm} $ is a normalization constant and 
\begin{eqnarray}
\Lambda_{\pm 1}^\dagger = \sum _m (m+1)^{\mp 1/2} d_{m+1,\downarrow} ^\dagger 
d_{m,\uparrow}. 
\end{eqnarray}
Adjacent regions correspond to values of $P$ which differ in one unit and, 
as a consequence, their spins also differ in 1.
Dashed lines in Fig. \ref{fig1} depict, for two different values of $\omega _0$, 
the evolution of the GS for a GaAs QD within the range of magnetic fields 
(perpendicular to the QD) given at the edges of the lines. 

Fig. \ref{fig2} shows the evolution of the GS properties from $\nu=2$ to 
$\nu =1$ for the system of Fig. \ref{fig1} with $\omega _0=
5.4$meV\cite{note2}. The GS energy $E_{GS}$ has a kink any time a crossing 
of states occurs. Since these kinks are not obvious in Fig.\ref{fig2}(b), 
we represent, in part (a), the energy splitting $\Delta E=E_{exc}-E_{GS}$ 
between the lowest excited state and the GS. The spin $S_z$ and the third 
component $M$ of the total angular momentum of the GS are also given in 
(c) and (d) respectively. 

Independently of the particular details, there is a general trend for 
any $N$: when the magnetic field increases from $\nu =2 $ 
to $\nu=1$, the GS changes many times (at least $N/2$ changes 
corresponding to electrons $N/2$ flipping their spins one by one). In other words, 
for $2>\nu>1$ there are several crossings of the lowest energy states. 
This means that, for many values of the field, the GS is doubly 
degenerated (or non degenerated but having an excited state with extremely low 
excitation energy). 
Moreover, in almost all the cases, these two states 
have spins differing in 1 while they have different $M$. 

\section{Kondo Hamiltonian}
As a consequence of the above numerical result, the main physics
of the problem is captured by a two level system approach \cite{Cox}.
One considers the QD
in the regime $2 > \nu > 1$ as having two degenerate (or almost degenerate) GS's 
$|GS_\Uparrow ^{N} \rangle$ and $|GS_\Downarrow ^{N} \rangle$ with $M_\Uparrow 
\neq M_\Downarrow $ and spins differing in 1, i.e.  $\langle GS_\Uparrow ^{N}
|S_z|GS_\Uparrow ^{N} \rangle= \langle GS_\Downarrow ^{N}|S_z|GS_\Downarrow ^{N} 
\rangle +1$. $H_{TUN}$ mixes these GS's with states $|N\pm1 \rangle $ 
in which $N\pm 1$ electrons are in the QD. In our two level system description, 
the tunneling Hamiltonian is projected on the subspaces subtended by the two 
GS's for $N$ electrons and the connecting $|N\pm1 \rangle$ electron states. 
In this process, we consider the connecting $|N\pm1 \rangle$ states as 
non-degenerate\cite{note3}.
Using the notation $\Sigma \equiv \{ \Uparrow , 
\Downarrow \}$, we introduce the tunneling spectral amplitudes  
\begin{eqnarray}
& & \Delta_{-,\Sigma}= \sum_{m,\sigma} V_m \langle N-1|d_{m,\sigma}|
GS ^{N}_{\Sigma}  \rangle 
\nonumber \\ 
& & \Delta_{+,\Sigma}= \sum_{m,\sigma} V_m \langle N+1
|d^\dagger_{m,\sigma}| GS ^{N}_{\Sigma}  \rangle \, 
\label{delta}
\end{eqnarray}
and the spin-flip Hubbard operators
\begin{eqnarray}
X_{\Sigma, \Sigma'}=|GS^N_\Sigma\rangle \langle GS^N_{\Sigma'} |.
\label{hubbards}
\end{eqnarray}
From projected tunneling Hamiltonian $\overline{H}_{TUN}$, the interaction 
between the QD with $N$ electrons and
the leads is studied by a standard scattering description \cite{Hewson}.
An effective coupling is obtained by summing up to all the possible 
intermediate states $|I\rangle$, in a second order perturbation approach: 
\widetext
\begin{eqnarray}
& & H_{eff}= \sum_{I} \frac{\overline{H}_{TUN}|I\rangle \langle I|
\overline{H}_{TUN}}{E_{GS}^N-E_I} 
\nonumber \\
& & = \sum_{k,k',\Sigma} \! \left[ 
\frac{\Delta_{-,\Sigma} ^2 \delta_{k,k'}X_{\Sigma,\Sigma}}
{E_{GS}^N- \! E^{N-1}- \! \varepsilon_F} 
+ \frac{\Delta_{+,\Sigma}^2c^\dagger_{k',\sigma}c_{k,\sigma}}
{E_{GS}^N- \! E^{N+1}+ \! \varepsilon_F} \!
+ \! \! \sum _{\Sigma'} \! \left(\frac{\Delta_{+,\Sigma}\Delta_{+,\Sigma'}}
{E^{N+1} \! - \! \varepsilon_F-\! E_{GS}^N}+
\frac{\Delta_{-,\Sigma}\Delta_{-,\Sigma'}}{E^{N-1}+
\! \varepsilon_F-\! E_{GS}^N} \! \right) \!  
X_{\Sigma,\Sigma'} c^\dagger _{k',\sigma'}c_{k,\sigma} 
\right] ,
\label{heff}
\end{eqnarray}
where, due to spin conservation, $\Sigma$ has the same direction than 
$\sigma$ and $\Sigma'$ the same than $\sigma'$.
The first term is simply a constant. The second term (involving $ 
c^\dagger_{k',\sigma}c_{k,\sigma}$) represents a potential scattering 
which does not involve any spin flip. These two terms are identical 
to the ones appearing when building up an $sd$ Hamiltonian from the 
Anderson Hamiltonian in the case of $N=1$\cite{Hewson}. 
As it is usually done in that case, one can forget 
about these two terms which do not contain anything important for the 
physics we want to address. 

\begin{figure}
\psfig{figure=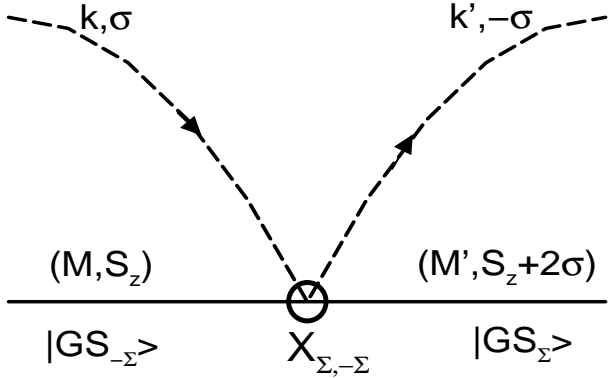,height=5.0cm,width=8.0cm}
\vspace{0.5cm}
\caption{Simplest diagram representing the spin-flip scattering described by the  
$X_{\Sigma,-\Sigma}=|GS_{\Sigma} ^{N} \rangle \langle GS_{-\Sigma} ^{N}|$ 
Hubbard operators. 
The base line depicts the evolution of the 
momentum and spin of the scatterer (the QD). Dashed lines depict carriers 
with spin $\sigma = \pm 1/2$ at the leads. Spin conservation requires that 
$\sigma $ has the same direction than $\Sigma $.} 
\label{fig3}
\end{figure}

The interesting physics is included in the third term of (\ref{heff}) 
which is a Kondo Hamiltonian
\begin{eqnarray}
& H_K= \sum_{k,k'} & \left[ 
J\left( X_{\Uparrow,\Downarrow} c^\dagger _{k'\downarrow}c_{k,\uparrow} 
+X_{\Downarrow,\Uparrow} c^\dagger _{k', \uparrow}c_{k,\downarrow} \right) 
+J_{\Uparrow} X_{\Uparrow,\Uparrow}c^\dagger _{k', \uparrow}c_{k,\uparrow}
+J_{\Downarrow} X_{\Downarrow,\Downarrow} c^\dagger _{k', \downarrow}c_{k,\downarrow} \right] .
\label{kondo}
\end{eqnarray}
with exchange couplings
\begin{eqnarray}
& & J \! = \! \frac{\Delta_{+,\Uparrow}\Delta_{+,\Downarrow}}
{E^{N+1}- \varepsilon_F-E_{GS}^N} \! +
\! \frac{\Delta_{-,\Uparrow}\Delta_{-,\Downarrow}}{E^{N-1}+
\varepsilon_F-E_{GS}^N} \, \, ; \, \, 
J_{\Sigma} \! =\! \frac{\Delta_{+,\Sigma}^2}{E^{N+1}-
\varepsilon_F-E_{GS}^N} \! + \! \frac{\Delta_{-,\Sigma}^2}{E^{N-1}+ 
\varepsilon_F-E_{GS}^N} \! .
\label{exchange}
\end{eqnarray} 
\narrowtext
$H_K$ is a spin-flip scattering Hamiltonian in which the GS
of the scatterer flips its spin by means of $X_{\Uparrow,\Downarrow}$ and 
$X_{\Downarrow,\Uparrow}$ (as shown in Fig. \ref{fig3}) and, at the same time, 
changes its total angular momentum. Both the difference between $M_\Uparrow $ 
and $M_\Downarrow$, and the correlation effects included in the tunneling 
spectral amplitudes, are the reason why one must use spin-flip Hubbard operators 
instead of the usual spin-ladder operators $S^{(\pm)}$. 

A crucial question is the sign of the exchange couplings (\ref{exchange}).
With our definitions (\ref{delta}), all the tunneling spectral amplitudes 
are positive.
Therefore, the signs of the exchange couplings are determined by the denominators.
As, in the considered situation, the lowest energy corresponds to having 
$N$ electrons in the QD, all the intermediate 
states $|I\rangle$, with $N \pm 1$ electrons in the QD and $\mp 1$ 
electron at the Fermi level of the leads, have higher energy. 
Therefore, we have a very 
important result: {\it $H_K$ has positive effective exchange couplings}. 
So, a Kondo Hamiltonian (\ref{kondo}) with {\it antiferromagnetic couplings} 
(\ref{exchange}) has been obtained. The conclusion is: 
{\it for any values of both $N$ and spin, the QD in the regime $2>\nu>1$ 
presents Kondo physics}. 

\section{Exchange couplings and Kondo temperature}
Since the Hamiltonian (\ref{kondo}) is formally equivalent to the standard 
Kondo Hamiltonian for $N=1$, the temperature dependence 
of the conductance (as any other measurable property) presents now  
characteristics similar to the one found in those experiments 
\cite{Goldhaber,Cronenwett,Simmel} which have been interpreted in terms
of only one electron. The important issue to be 
discussed is the characteristic energy scale, $T_K$, which 
is determined by the antiferromagnetic couplings. 

$J$ and $J_{\Sigma}$ depend on both the energy difference 
$E_{GS}^{N}-E^{N \pm 1}\mp \varepsilon_F$, 
and tunneling spectral amplitudes $\Delta_{-,\Sigma}$. The 
former is practically independent on both $N$ and $\Sigma$. 
Therefore, it does not imply any significant difference with respect to the 
well known case of $N=1$. However interesting physical differences appear 
due to $\Delta _{-,\Sigma}$. Let us analyze the different situations in which 
the QD can be found:

{\it i)} (Upper part of Fig. \ref{fig1}) 
When the Zeeman interaction is large (for instance having a 
large inplane component of the magnetic field), the GS's are always 
compact states 
\begin{eqnarray}
|GS_\Uparrow ^{N} \rangle=|C^K_{N-K} \rangle  \, \, \, ; \, \, 
|GS_\Downarrow ^{N} \rangle=|C^{K+1}_{N-K-1}\rangle .
\end{eqnarray}
This is the simplest case in which 
\begin{eqnarray}
X_{\Uparrow,\Downarrow}=(-1)^K d^\dagger _ {N-K-1, \uparrow}d_ {K, \downarrow}
\end{eqnarray}
and similarly for the other Hubbard operators.
We have also assumed, $|N-1\rangle= 
|C^{K}_{N-K-1} \rangle$ and $|N+1\rangle=|C^{K+1}_{N-K} \rangle$
(in other case the quantum numbers of the 
$d^\dagger$ operators must be changed accordingly) and the signs are taken 
according with definitions (\ref{hubbards}). 
Due to the lack of correlation effects in the GS's, the tunneling spectral 
amplitudes are 
\begin{eqnarray}
& & \Delta_{-,\Uparrow}= \Delta_{+,\Downarrow}=V_{N-K-1} \, ; \, 
\Delta_{-,\Downarrow}=\Delta_{+,\Uparrow}=V_{K}.
\end{eqnarray}
There is a difference between the antiferromagnetic 
couplings $J$, $J_\Uparrow$ and $J_\Downarrow$ due to the fact that $N-K-1 
> K$. It is easier to tunnel from the leads to spin up states 
($m=N-K-1$) within
the QD because they are in the outer region of the QD while the first 
available spin down state ($m=K$) is in the inner region of the QD. 
This tunneling amplitude effect is not described by previous models
of Kondo at finite B \cite{Pustilnik,Eto,Giuliano}, but has long been 
more broadly recognized both experimentally and theoretically for 
QD with $2>\nu>1$.

Although in this regime the QD does not present any correlation effects, 
the antiferromagnetic couplings are still a function of total energies and 
tunneling amplitudes. Since the Kondo temperature depends exponentially on 
$J$\cite{Hewson}, it is a very sensitive magnitude with respect to many 
parameters as $w$, $\omega _0$, $B$, $V_m$ etc. However, this is not 
different from the case of just one electron in the QD in which experiments 
\cite{Goldhaber,Cronenwett,Simmel} show $T_K$ to be in the range of 1K.
In any case, one can predict variations in $T_K$ when 
the regime changes as discussed below. 

{\it ii)} (Lower part of Fig. \ref{fig1}) 
When the Zeeman interaction is small enough, the GS's are skyrmion-like states 
of topological charge one \cite{Oaknin1,Oaknin2}:
\begin{eqnarray}
|GS_\Uparrow ^{N} \rangle=|SK^P_{N,K,\pm} \rangle \, ; \, 
|GS_\Downarrow ^{N} \rangle=|SK ^{P+1}_{N,K,\pm}\rangle 
\label{skyrmions}
\end{eqnarray}
The Hubbard operators $X_{\Sigma, \Sigma '}$ have 
complicated, but analytical, expressions in terms of $\Lambda_{\pm 1}^\dagger $.

When $N$ is very large, correlation effects provoke that $\Delta _{-,\Sigma}$ 
tends to zero (for instance 
$\Delta _{-,\Sigma} \propto [N ln N]^{-1/2}$ for $P=0$) 
implying a quenching of the antiferromagnetic couplings due to the 
orthogonalization catastrophe. As a consequence, Kondo effect should not 
be observed when the number of electrons within the QD is large because 
the Kondo temperature is extremely small in this case. 

Even more interesting is the effect for a reduced number 
of electrons, for instance $N=8$ as both in Fig. \ref{fig1}, \ref{fig2} and in 
the experiment \cite{Sttut}. In this case,  
correlation effects do not destroy Kondo effect but reduce $\Delta_{-,\Sigma}$ 
up to a factor of two \cite{note1}. The reduction of $J$ and $J_{\Sigma}$ 
implies that Kondo temperatures are significantly smaller 
than for the case {\it i)} of compact states. In practice, one can move from 
the skyrmions regime {\it ii)} (lower part of Fig \ref{fig1}) to the compact 
regime {\it i)} (upper part of Fig \ref{fig1}) by increasing an inplane 
component of the magnetic field. As a consequence of the above analysis, 
one should detect a clearly measurable increase of $T_K$ during this process.

{\it iii)} There is a rather unusual case, in which, for intermediate Zeeman 
interaction and only for a few values of the magnetic field, the degenerate 
GS's are one skyrmion and one compact state, i. e.:
\begin{eqnarray}
|GS_\Uparrow ^{N} \rangle= |SK^P_{N,K,+}\rangle 
\, ; \,  |GS_\Downarrow ^{N} \rangle=|C^{K-1}_{N-K+1}\rangle .
\end{eqnarray}
This happens, for instance, in the last step to the right in Fig. \ref{fig2}. 
In this case, the two GS have spins differing in more than 1. Therefore, 
they can not be obtained from the same $|N-1 \rangle $ by creating  
one electron with spin either up or down.
In other words, one of the two GS's has a tunneling spectral amplitude equal 
to zero so that, for tunneling effects there is only a non degenerate GS 
and Kondo effect does not occur. 

A practical difference between the compact and skyrmion cases resides in the 
probability of finding two degenerate (or almost) GS's. As 
shown in Fig. \ref{fig1}, the crossing regions are much closer to each other 
for skyrmions (lower part of Fig. 
\ref{fig1}) than for compact states (upper Fig. \ref{fig1}).
To work with a magnetic field in which Kondo effect exists 
is more probable in the skyrmion regime than in the compact one.
However, once Kondo regime is found in the compact regime, the effect is 
more robust because its $T_K$ is higher than the one for skyrmions.

\section{{\it Chess board} behavior of the conductance}

A very characteristic feature of some experiments \cite{Sttut,Delft2} is 
an alternating high-low conductance sequence as a function of $B$ for a 
given temperature and number of electrons. When $N$ is varied in $\pm 1$, 
the high and low conductance valleys are interchanged. Since a
common way of representing the experimental results is to use a color-intensity 
scale for the conductance as a function of both $B$ and a gate potential which 
varies $N$, the data present a {\it chess board} aspect\cite{Sttut,Delft2}.
This occurs in a broad range of filling factors and number of electrons.   

In order to understand such an experimental behavior, let us start by 
applying our description for the regime $1<\nu<2$. Since the {\it chess board} 
is observed \cite{Sttut,Delft2} in samples with large $w$, skyrmion-like
states are high energy excitations. GS's are restricted to be 
given just by one configuration, i. e. $[S^{(-)}]^j|C^K_{N-K} \rangle$ 
states, where $j$ is an integer (for practical purposes one can 
concentrate in the cases $j=0$ and $1$).
At first sight, the consideration of both $|C^K_{N-K} \rangle$ and 
$S^{(-)}|C^K_{N-K} \rangle$ seems unnecessary because these 
states have an energy difference equal to the Zeeman coupling so that they 
are never degenerate. However, in 
a broad range of magnetic fields, these two states are precisely the two 
lowest energy states (see e.g. Fig. \ref{fig2}).

\begin{figure}
\psfig{figure=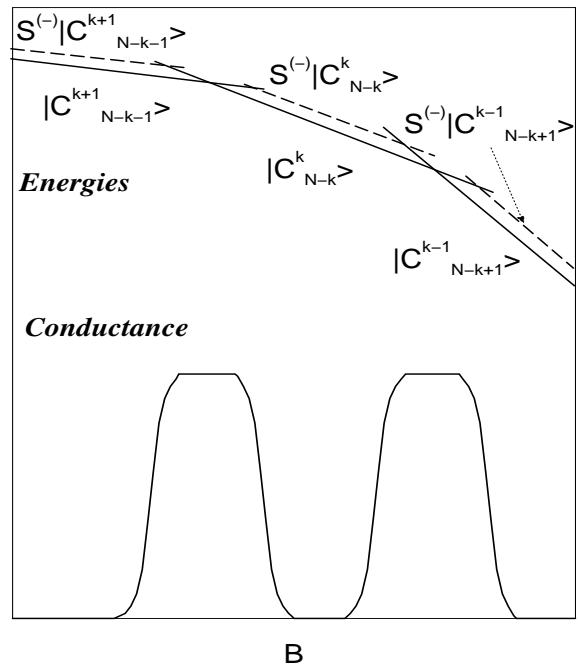,height=9.0cm,width=8.5cm}
\caption{Schematic evolution with the magnetic field of the two lowest 
energy states of a QD (upper part) and the conductance (lower part) in the 
range $T_K > T > \Delta E^2/T_K$ (see text).}
\label{fig4}
\end{figure}

In the two level system approach\cite{Cox} in which the two lowest states
are separated by an energy splitting $\Delta E$, Kondo-like  
behavior appears only when the experimental temperature is between
$(\Delta E)^2/T_K$ and $T_K$ (provided $\Delta E < T_K$) 
in order to have significant occupation of the two states. 
The quantitative application of the model would require the computation of 
both the energy splitting $\Delta E$ and $T_K$ which, as mentioned 
above, are very sensitive to many experimental parameters. Instead, a 
general understanding of the {\it chess board} is obtained from the following
qualitative explanation depicted schematically in Fig. \ref{fig4}:
when magnetic field is varied, the two lowest levels are in two alternating 
situations depending whether the system is around or far from a
crossing of GS's. Around a crossing, 
the two lowest levels correspond to states $|C^{(K+1)}_{N-K-1} \rangle$ and  
$|C^K_{N-K}\rangle$ while far from the crossing they are $|C^K_{N-K}\rangle$ 
and $S^{(-)}|C^K_{N-K} \rangle$. In the former case, the two lowest levels 
can be connected by the tunneling Hamiltonian through low energy states 
$|N\pm 1\rangle $ with $\Delta_{\pm, \Sigma}=V_KV_{N-K-1}$. In this case,
the system presents high Kondo-like conductance. In the second case of being 
far from the crossings, the two lowest states are $|C^K_{N-K}\rangle$ and 
$S^{(-)}|C^K_{N-K} \rangle$. The coupling $J$ between the two states is 
very weak for two reasons: first, the connecting states $|N\pm 1\rangle $ are 
highly excited states implying very large denominators in $J$ and, second, 
since $S^{(-)}|C^K_{N-K} \rangle$ is a linear combination of different 
configurations, $\Delta _{\pm,\Sigma}$ is significantly reduced 
by factors proportional to $1/N$. These two effects produce, in this region,
a very small $J$ and, consequently, an exponentially negligible $T_K$ so that 
the measured conductance is not Kondo-like but instead is quenched. This 
originates the alternating behavior experimentally observed for fix gate 
voltage (i. e. number of electrons in the QD) and varying magnetic field.
The {\it chess board} aspect also implies alternating low-high conductance  
regions for fix $B$ and varying gate voltage, which is due
to the fact that crossings for $N\pm 1$ electrons occur for magnetic fields 
roughly midway from crossings for $N$ electrons\cite{note1}.

Since Kondo-like behavior only occurs when the experimental temperature 
is in the range between $(\Delta E)^2/T_K$ and $T_K$, 
a clear prediction of our scheme is that the highly conducting regions of the 
{\it chess board} would become narrower for decreasing temperature due to the 
lower limit condition.

The sequence of alternating high-low conductance has been observed also 
at very low magnetic field ($B\simeq 0.5$T) and large number ($N\simeq 50$) 
of electrons\cite{Delft2}. 
So, a complete understanding of the experimental situation requires the 
extension of our framework to $\nu >2$. We are currently involved in this 
task which implies to take into account GS's more complicated than  
simple compact states in the lowest Landau level. 

\section{Summary}
We have presented a general description of the Kondo effect
for any number of electrons in a QD at filling factor $1<\nu <2$.
Collective spin effects of the $N$ electrons are described 
in terms of a set of spin-flip Hubbard operators 
which also change the third component of the total angular momentum.
Our conclusion is that, for any number of electrons within the QD, 
the spin-flip scattering of carriers only produces transitions between two 
(either compact or skyrmion-like) states of the QD with spins differing in 1. 
This process can be described by a Kondo Hamiltonian with exchange couplings
that, we show, are antiferromagnetic and depend on both correlation and 
tunneling amplitude effects. As a consequence, the system presents 
a Kondo behavior with an experimentally accessible $T_K$. 
The increase of $T_K$  with an inplane component of the magnetic field 
should allow the detection of the transition from a skyrmion-like 
regime to a compact state regime. Finally, we present a qualitative 
explanation for the {\it chess board} aspect of the experimental conductance 
when represented in a grey scale as a function of both the magnetic field 
and the gate potential affecting the quantum dot. 

\section{Acknowledgements}
We are grateful to J. Weis and L. P. Kouwenhoven for providing us with 
experimental information before its publication.
This work was supported in part by MEC of Spain under contract No. PB96-0085,
Fundacion Ramon Areces and CAM under contract No. 07N/0026/1998.

\widetext


\begin{references}
\bibitem{note3} Non-degeneracy of $|N\pm1\rangle $ states is the most common 
case. If either $|N+1\rangle $ or $|N-1\rangle $ is also 
degenerate, the algebra is more complicated but the physics is the same.
\end{references}
\end{document}